# Maladaptation and the Paradox of Robustness in Evolution

Steven A. Frank*

Department of Ecology and Evolutionary Biology, University of California at Irvine, Irvine, California, United States of America

*Background.* Organisms use a variety of mechanisms to protect themselves against perturbations. For example, repair mechanisms fix damage, feedback loops keep homeostatic systems at their setpoints, and biochemical filters distinguish signal from noise. Such buffering mechanisms are often discussed in terms of robustness, which may be measured by reduced sensitivity of performance to perturbations. *Methodology/Principal Findings.* I use a mathematical model to analyze the evolutionary dynamics of robustness in order to understand aspects of organismal design by natural selection. I focus on two characters: one character performs an adaptive task; the other character buffers the performance of the first character against perturbations. Increased perturbations favor enhanced buffering and robustness, which in turn decreases sensitivity and reduces the intensity of natural selection on the adaptive character. Reduced selective pressure on the adaptive character often leads to a less costly, lower performance trait. *Conclusions/Significance.* The paradox of robustness arises from evolutionary dynamics: enhanced robustness causes an evolutionary reduction in the adaptive performance of the target character, leading to a degree of maladaptation compared to what could be achieved by natural selection in the absence of robustness mechanisms. Over evolutionary time, buffering traits may become layered on top of each other, while the underlying adaptive traits become replaced by cheaper, lower performance components. The paradox of robustness has widespread implications for understanding organismal design.

## INTRODUCTION

I argue that robustness creates a directionality in the evolutionary history of life. Each increase in robustness reduces the sensitivity of certain organismal traits to environmental perturbations [1–5]. Reduced sensitivity often means that degradation in the trait has less consequence for organismal fitness. If a degraded trait does not reduce fitness as much as it would have before protection by a robustness mechanism, then natural selection will often favor a less costly, lower performance trait. Over time, the dynamics play out as follows: robustness mechanisms reduce sensitivity; those traits buffered by robustness degrade to lower performing and less costly states; new robustness mechanisms arise; and the cycle continues. Buffering traits become layered on top of each other, while the underlying traits become replaced by cheaper, lower performance components.

To fix ideas, I begin with a particular example and then follow with a general formulation of the robustness paradox. In the introductory example, the adaptive character is a receptor that distinguishes between correct and incorrect ligands. The level of perturbation arises from the cost of incorrect binding. Misbinding depends on the abundance of alternative ligands, which sets the level of background noise from which the receptor must distinguish the correct signal. To study this problem, I follow the classic model of kinetic proofreading [6] as formulated by Alon [5].

### Relation to past research

Studies of robustness have followed distinct lines of thought and distinct literatures. Before proceeding with my own analyses, it is useful to place my approach within the context of these existing lines of thought.

Recently, the most widely discussed research arises from the study of genetic robustness: the mechanisms that reduce the sensitivity of organismal performance to inherited mutations [1,4]. That research typically considers how organisms evolve to protect themselves against perturbations from the environment, perturbations during development, or perturbations from inherited mutations. Buffering mechanisms that protect against those various kinds of perturbations have the consequence of reducing the sensitivity of the organism to an inherited mutation, even if the original buffering mechanism arose to protect against environmental or developmental perturbations. Reduced sensitivity to inherited mutations slows the rate at which natural selection clears deleterious mutations from the population. Slower clearance of deleterious mutations increases the accumulation over time of genetic variability.

This view of genetic robustness plays a key role in understanding the evolutionary forces that shape genetic variation. In this case, the particular argument depends on the balance between natural selection and mutation: on one side, robustness reduces sensitivity to genetic (mutational) perturbations and therefore weakens the force of natural selection; on the other side, the weakened force of selection must be balanced against the continual evolutionary pressure from new mutations.

This argument about the balance between mutation and selection may be important for understanding genetic variation. But it does not help in understanding the evolutionary design of organismal characters. The problem is that mutation is a relatively

. . . . . . . . . . . . . . . . . . . . . . . . . . . . . . . . . . . . . . . . . . . . . . . . . . . .

Funding: National Science Foundation grant DEB-0089741 and National Institute of General Medical Sciences MIDAS Program grant U01-GM-76499 support the research of this author.

Competing Interests: The author has declared that no competing interests exist.

* To whom correspondence should be addressed. E-mail: safrank@uci.edu

weak evolutionary force, so changes in the balance between mutation and selection usually do not cause major changes in the design of the key performance characteristics of organisms.

By contrast, a completely distinct line of thought and literature focuses on the costs and benefits of key performance characteristics in the face of environmental perturbations [7–9]. This literature emphasizes safety factors in design, that is, excess capacity of characters that protect against failure in the face of rare perturbations. Safety factors are just another name for buffering or robustness, in that such factors reduce the sensitivity of the organism to perturbations. The literature on safety factors emphasizes a cost-benefit analysis. Safety factors are costly, and so must provide sufficient benefit in protecting against perturbations in order to offset their costs. By such cost-benefit analysis, one may analyze the variability in safety factors found in different characters, such as bones or respiratory capacity of lungs.

I initiate in this paper a different argument that derives aspects from the two prior traditions. On the one hand, I focus on how costs and benefits shape key performance characters, like the work on safety factors. On the other hand, I emphasize the dynamics of evolutionary change, as in the literature on genetic robustness. Joining cost-benefit analysis with evolutionary dynamics leads me to study how key performance characters change evolutionarily in response to changes in robustness or, similarly, to changes in safety factors.

## Overview of the argument

The broad subjects of robustness and safety factors cover many different kinds of problems. I focus narrowly on my single point: the cost-benefit analysis of performance traits leads to interesting conclusions about the joint evolution of robustness and decay.

I begin with the explicit example of kinetic proofreading to clarify key aspects of my argument. In particular, the example shows how I separate between traits that buffer the organism from perturbations and traits that directly perform an important function. Increased buffering alters the balance between costs and benefits for the direct performance character. I study the joint evolution of the two characters: buffering and direct performance. The evolutionary dynamics of these two characters leads to my conclusions about how increases in robustness may lead to the decay or maladaptation of direct performance characters.

After I develop the particular example to illustrate my key points, I then turn to a general analysis of the problem. The general analysis provides insight into any system that can be understood in terms of distinct buffering and direct performance characters. Along the way, I derive measures for robustness and maladaptation. Those measures clarify how to think about the evolutionary dynamics.

I develop measures of robustness and maladaptation in terms of the shapes of fitness surfaces. That approach shares certain features with prior work, for example, that by Rice [10]. The common use of such fitness surfaces is not surprising. Ultimately, to understand evolutionary dynamics, one must implicitly or explicitly be tracking how changes in characters alter fitness surfaces and, in turn, how altered fitness surfaces influence the evolution of characters.

## RESULTS

### Cost-benefit analysis: an example

In this biochemical example, the receptor binding kinetics of correct and incorrect ligands differ only in their off-rates, the rates at which the bound ligands release from the receptor. The correct ligand releases at rate $k$; the incorrect ligands release at rate $k(1+\delta)$. Kinetic proofreading arises from the delay time, $\tau$, between initial binding and when the bound ligand induces a signal in the receptor. With this setup, the probability that a correct initial binding yields a signal is $e^{-k\tau}$, and the probability that an incorrect initial binding yields a signal is $e^{-k(1+\delta)\tau}$. The ratio of correct to incorrect signals is $e^{\delta k\tau}$. An increased delay, $\tau$, enhances the ratio of accurate to inaccurate signals. Note that $k\tau$ always occurs as a pair, so, without loss of generality, I set $k=1$ and absorb that parameter into values of $\tau$.

This model of kinetic proofreading has provided much insight into several aspects of molecular recognition [5,11,12]. However, the simple ratio of correct to incorrect signals, $e^{\delta\tau}$, is incomplete from an evolutionary perspective. The ratio depends on two characters: $\delta$ sets the intrinsic discriminatory power of the receptor to distinguish correct and incorrect ligands; $\tau$ sets the strength of the filter that separates correct recognition from perturbations that increase with the abundance of incorrect ligands. Increases in either the direct adaptive character, $\delta$, or in the buffer against perturbation, $\tau$, raise the success ratio. Certainly, natural selection will not increase both characters without bound. So, how does natural selection jointly shape the direct adaptive character and the buffer against perturbation?

An increase in the buffering character, $\tau$, lowers the probability that any ligand induces a signal. Thus, a rise in $\tau$ reduces the correct signals as well as the incorrect signals. The benefit arises because the incorrect signals are reduced more than the correct signals, but a full analysis must also account for how increases in $\tau$ reduce the correct signal probability. In addition, the simple formulation does not provide any limit on how $\delta$ will evolve: without any explicit constraint, the receptor would evolve ever increasing discriminatory power by raising $\delta$ until incorrect ligands fell off at very high rates.

We can account for all costs and benefits on both characters by formulating fitness, or performance, as

$$w = e^{-c\delta}e^{-\tau}(1-be^{-(1+\delta)\tau}), \qquad (1)$$

where $e^{-c\delta}$ is the cost imposed on receptor function for an increased ability to discriminate against incorrect ligands; $e^{-\tau}$ is the probability of a correct ligand signal per correct ligand binding; and $be^{-(1+\delta)\tau}$ is the probability of an incorrect ligand signal per incorrect ligand binding weighted by the intensity and cost of such perturbations, $b$. Thus, we have two character values, $\delta$ for direct discrimination against incorrect signals and $\tau$ as a buffer that reduces the sensitivity to incorrect signals. Two parameters influence how those characters determine fitness: $c$ sets the cost of direct discrimination, and $b$ sets the intensity of external perturbations.

The solid lines in Figure 1A–F show the optimum values for the buffer character, $\tau^*$, and the direct adaptive character, $\delta^*$, for different intensities of perturbation, $b$, and costs for the direct character, $c$. To obtain the joint optimum, $\{\delta^*,\tau^*\}$, I let both characters evolve in response to each other.

How much does the buffer character reduce the performance of the direct character by shielding the direct character from the consequences of perturbations? To study that question, I set the buffer character to a fixed value, $\hat{\tau}$, the optimum value of $\tau$ when $b=1$. I then increased $b$, forcing the direct character, $\delta$, to respond fully to the rising intensity of perturbations without the benefit of being protected by enhanced buffering.

The dashed lines in Figure 1D–F show the optimum value of the direct character, $\tilde{\delta}$, given a fixed buffering character, $\hat{\tau}$. In the absence of protection by enhanced buffering, the direct character evolves to significantly higher levels of performance in response to rising perturbations. The decline of the solid lines relative to the dashed lines shows the degree of reduced performance for the direct character caused by enhanced buffering. To measure reduced performance of the direct character, $\delta$, note that $e^{-(1+\delta)\tau}$ measures the probability of a false signal caused by binding an incorrect ligand. One way to measure the decline in performance between $\tilde{\delta}$ (dashed lines) and $\delta^*$ (solid lines) in Figure 1D–F, holding other factors constant, is by the logarithm of the ratio of false signal probabilities given a fixed value of $\tau$,

$$\log\left(\frac{e^{-(1+\delta^*)\tau}}{e^{-(1+\tilde{\delta})\tau}}\right) = (\tilde{\delta} - \delta^*)\tau \qquad (2)$$

thus, the distance between the dashed and solid lines provides a reasonable measure of the reduced performance of the direct character caused by an evolutionary response to buffering. Put another way, the robustness provided by the evolution of the buffering character, $\tau$, leads to maladaptation in the character that directly affects the phenotype under study—in this case, the filtering of false signals.

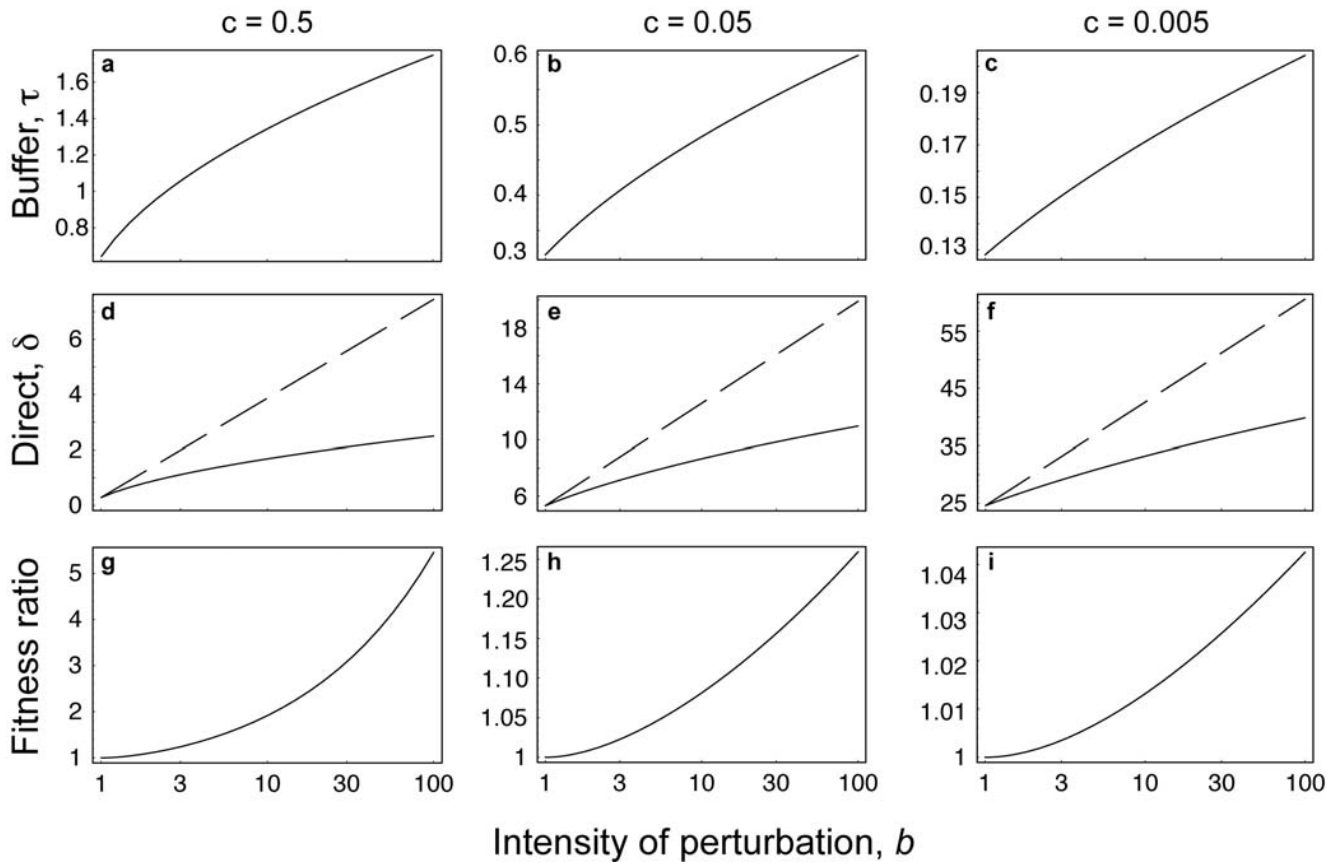

**Figure 1. Joint evolution of a direct character, $\delta$, that filters signal from noise, and a buffering character, $\tau$, that reduces the sensitivity of the direct character to perturbations.** The intensity of perturbations rises with the cost of incorrect signals, $b$. The direct character has an independent cost, $c$. (A–C) The optimal value of the buffering character, $\tau^*$, when both the buffering character and the direct character evolve freely. (D–F) The optimal value of the direct character. The solid line shows $\delta^*$, when both the direct and buffering character evolve freely. The dashed line shows $\tilde{\delta}$, the evolutionary response of the direct character when the buffering character is constrained to $\hat{\tau}$, as explained in the text. (G–I) The ratio of fitness, $w$, when both characters evolve freely to $\{\delta^*, \tau^*\}$ relative to when $\tau$ is fixed and $\delta$ evolves freely to the point $\{\tilde{\delta}, \hat{\tau}\}$. The ratio shows the net benefit of robustness in terms of enhanced fitness from buffering against perturbations.
doi:10.1371/journal.pone.0001021.g001

## The meaning of maladaptation

By *maladaptation*, I mean the decrease in the performance of a character relative to what could be achieved by natural selection in the absence of robustness mechanisms. For example, corrective lenses for vision, by buffering against genetic and developmental perturbations, may reduce native visual acuity over evolutionary time, even though net acuity—native acuity plus corrective lenses—may increase. In that regard, I consider the loss of native acuity to be maladaptation. It would be more precise to say that visual acuity suffers partial maladaptation, holding constant the buffering mechanism of corrective lenses. Because we often evaluate direct performance, holding other characters constant, it is useful to consider the reduced direct performance that results from the introduction of a distinct character that acts as a buffer against perturbations.

## Alternative ways to view robustness

I now turn to the problem of measuring robustness. Robustness means reduced sensitivity to perturbation. Reduced sensitivity typically acts by a buffer that lowers the intensity of natural selection on the direct character; the lower intensity of selection is what leads to a decay in performance of the direct character. In the particular example here, how does the buffering character, $\tau$, reduce sensitivity of the direct character, $\delta$?

One way to measure reduced sensitivity is to start at a particular optimum, $\{\delta^*, \tau^*\}$, and then analyze how a small increase in $\tau$ reduces the intensity of natural selection on $\delta$. This measure provides insight into a necessary condition for $\tau$ to be considered a robustness mechanism: greater buffering must reduce sensitivity of some direct character. I discuss this measure below.

A second measure emphasizes the evolutionary dynamics in response to increased perturbations and the consequences for the accumulation of deleterious genetic variation by mutation pressure. Suppose, for some level of perturbation, $b_1$, we are at an optimum $\{\delta_1^*, \tau_1^* | b_1\}$. Perturbations subsequently increase to $b_2$, leading eventually to a new optimum $\{\delta_2^*, \tau_2^* | b_2\}$, where, in response to greater perturbations, buffering has increased, $\tau_2^* > \tau_1^*$. With increased buffering under perturbation conditions $b_2$, how has the sensitivity of $\delta$ changed? In particular, how does a given fractional change in $\delta$ affect fitness under the stronger buffering condition $b_2$ relative to the weaker buffering condition $b_1$?

## The paradox of robustness in the kinetic proofreading example

For the particular model of kinetic proofreading in Equation 1, Figure 2 shows the reduction in sensitivity to fluctuations in $\delta$ under the stronger buffering condition. Here, reduced sensitivity means a smaller effect on fitness for a given percentage change in $\delta$. With reduced sensitivity, the intensity of natural selection has declined, and thus mutations that cause deviations in the direct character, $\delta$, will spread more easily, causing decay in the direct performance of the adaptive character. Such decay further exacerbates the maladaptation caused by robustness.

This example of kinetic proofreading illustrates the paradox of robustness. As mechanisms that increase robustness spread, the system gains increasing fitness by reduced sensitivity to perturbations. However, reduced sensitivity causes the performance of direct characters to decay, leading to greater maladaptation of the direct characters when measured by holding constant other characters. The decay arises from two processes. First, greater buffering reduces the benefit provided by the direct character, favoring the use of less costly components or cheaper designs. Second, greater buffering, by reducing the intensity of natural selection acting on the direct characters, allows greater accumulation of mutations that decay performance of the direct characters. In general, mutation is a relatively weak force. Thus, the first process, in which natural selection favors the use of cheaper designs, will usually dominate the second process, in which mutation accumulation decays performance.

I have, thus far, presented all of my analyses and quantitative measures in terms of the particular example of kinetic proofreading. That example fixed ideas and showed how one must analyze the joint evolution of two characters to understand the evolutionary dynamics of robustness. I now turn to a more general quantitative analysis that facilitates application to other problems.

## General measures of robustness

For a buffering character to qualify as a robustness mechanism, it must reduce the sensitivity to perturbations of a direct character. Denote, as above, a buffering character, $\tau$, a direct character, $\delta$, and the intensity of perturbation, $b$. Here, these values are not tied to a particular model, but represent the general expression of the problem. Let system performance, or fitness, be $w(\delta,\tau,b)$. For $b$ to act as a perturbation, an increase in $b$ must reduce $w$, that is, $\partial w/\partial b < 0$. For $\tau$ to act as a buffer against perturbation, an increase in $\tau$ must reduce the negative effect of increasing perturbation,

$$\frac{\partial}{\partial \tau}\left(\frac{\partial w}{\partial b}\right) > 0.$$

At a local optimum, $\{\delta^*, \tau^* | b^*\}$, local sensitivity of fitness to changes in the direct character can be measured by the curvature of the fitness function at the optimum

$$\sigma = \frac{\partial^2 w/w^*}{(\partial \delta/\delta^*)^2} < 0,$$

where the value of $\sigma$ is negative because the second derivative decreases around a local optimum. The buffer character reduces local sensitivity of the direct character by an amount measured by the degree to which an increase in buffering, $\tau$, reduces the curvature of the fitness function in relation to the direct character: reduced curvature means that $\sigma$ becomes less negative (closer to zero), causing a flattening of the fitness curve near the optimum. This reduction in curvature, a measure of robustness, can be expressed by

$$\rho = \frac{\partial \sigma}{\partial \tau/\tau^*} > 0.$$

These definitions of local sensitivity define robustness induced by the character $\tau$ in relation to fitness and in relation to a direct character, $\delta$. One may often wish to have a global measure of changes in sensitivity of the direct character to changes in $\tau$, in order to predict the amount of additional genetic variation that may accumulate by mutation in response to reduced selective intensity on $\delta$. Suppose we wish to compare the system at two different points. To obtain the first point, which we use for reference values, we first calculate the optimum at $\{\hat{\delta}, \hat{\tau} | \hat{b}\}$. We then calculate the constrained optima $\{\tilde{\delta}, \hat{\tau} | b\}$ for $b > \hat{b}$, in which we keep $\tau = \hat{\tau}$ and thereby force all changes in response to increased perturbations to be made via the direct character, $\delta$. This setup provides reference values for the system when it is forced to adjust to increased perturbations without the benefit of increasing robustness via the buffering character, $\tau$. We compare these reference points, with $\tau$ constrained not change, against optima that result when the system is allowed to enhance buffering in response to increased perturbation. In particular, we start at the same point, $\{\delta^*, \tau^* | b^*\} = \{\hat{\delta}, \hat{\tau} | \hat{b}\}$, but as $b$ increases, we allow both $\delta^*$ and $\tau^*$ to change, in general, causing $\tau^* > \hat{\tau}$ and $\delta^* < \tilde{\delta}$.

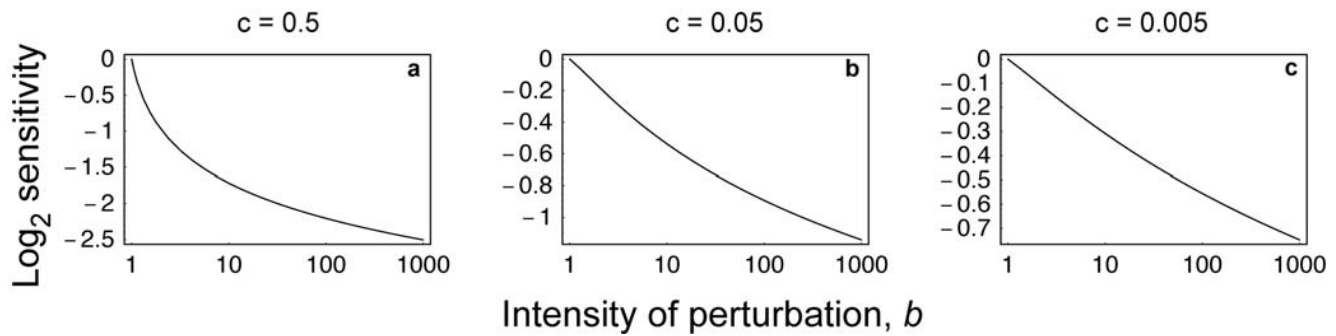

**Figure 2. Reduced sensitivity caused by robustness that buffers against perturbations.** Calculated as $\log_2(\gamma)$, where $\gamma$ is defined in Equation 3, and $\varepsilon = 0.01$.
doi:10.1371/journal.pone.0001021.g002

With these definitions, we can develop a global measure for the change in the sensitivity of the direct character, $\delta$, to a change in $\tau$, in order to predict the amount of additional genetic variation that may accumulate by mutation in response to reduced selective intensity on $\delta$. We compare the system at the two different optima described in the previous paragraph. In this case, we measure global sensitivity, $\gamma$, by the change in relative fitness for a fractional change, $\varepsilon$, in the optimum of the direct character, $\delta_{opt}$, that is, $\delta = \delta_{opt}(1-\varepsilon)$, yielding

$$\gamma = \frac{w(\delta^*(1-\varepsilon), \tau^*, b)/w(\delta^*, \tau^*, b)}{w(\tilde{\delta}(1-\varepsilon), \hat{\tau}, b)/w(\tilde{\delta}, \hat{\tau}, b)} \quad (3)$$

In general, $\gamma < 1$ when $\tau$ acts as a robustness mechanism that reduces the sensitivity of performance to fluctuations in $\delta$. Figure 2 plots $\log_2(\gamma)$ for the particular example of kinetic proofreading.

## A general measure of maladaptation shaped by cost and benefit

Finally, we need a measure of maladaptation that describes the partial decay in performance caused by evolution of the direct character to a lower cost solution in response to the protection provided by robustness mechanisms. As with sensitivity, I begin with a local measure, then develop a global measure. For the local measure, I start at a local optimum $\{\delta^*, \tau^*\}$, and use the chain rule to partition how $\delta^*$ changes in response to an increase in perturbation

$$\frac{d\delta^*}{db} = \frac{\partial \delta^*}{\partial b} + \frac{\partial \delta^*}{\partial \tau^*} \frac{d\tau^*}{db}.$$

In general, a robustness mechanism implies that buffering increases in response to an increase in perturbations, $d\tau^*/db > 0$, and an increase in buffering causes a partial reduction in the direct character by the protection afforded against perturbations, $\partial \delta^*/\partial \tau^* < 0$. Thus, the response of the buffering character, $\tau$, causes $\delta^*$ to increase in response to enhanced perturbations by less than it would in the absence of buffering—the term

$$\lambda = \frac{\partial \delta^*}{\partial \tau^*} \frac{d\tau^*}{db} < 0$$

measures the local pressure causing maladaptation of $\delta$.

A global measure of maladaptation requires comparison of two local optima. For one of the optima, we constrain $\tau$ to be unchanging in response to perturbation, forming a reference point against which we can compare the response of $\delta$ when $\tau$ is not constrained. As before, we begin at the point $\{\delta^*, \tau^* | b^*\} = \{\hat{\delta}, \hat{\tau} | \hat{b}\}$. As $b$ increases, we constrain one point to keep a fixed value of $\tau = \hat{\tau}$, yielding the reference optimum $\{\tilde{\delta}, \hat{\tau} | b\}$, and compare that optimum to the one that arises when both $\delta$ and $\tau$ evolve jointly to the point $\{\delta^*, \tau^* | b\}$. A global measure of maladaptation compares the partial direct performance of $\delta$ in the absence of enhanced buffering compared with enhanced buffering, at a value $b > \hat{b}$, yielding

$$\mu = P(\tilde{\delta}) - P(\delta^*),$$

where $P(\cdot)$ measures partial direct performance, holding constant other factors. The best definition of $P(\cdot)$ may depend on the particular problem and on the factors one wishes to emphasize. In my example of kinetic proofreading, the logarithm of the ratio of false signal probabilities given a fixed value of $\tau$ was given by Equation 2: in that particular example $P(\delta) = \delta\tau$ is a linear function of the direct character, $\delta$.

## DISCUSSION

The paradox of robustness arises because each increase in buffering causes the evolution of reduced performance by direct adaptive characters. This feedback between robustness and the direct characters that are protected from perturbations leads to a directionality in evolution: enhanced robustness protects against perturbations and increases fitness; direct characters decay because they are shielded from perturbations; enhanced robustness, once in place, cannot easily be removed, because the decay of the original direct characters causes the system to depend on the new protections against perturbations. In consequence, additional robustness mechanisms will, over time, be layered on top of the system; direct characters will decay; and new protections against perturbations will follow. Other factors will, of course, come into play, but evolutionary dynamics drives systems in the direction of repeated rounds of enhanced robustness and decay.

The fundamental logic of this theory holds without doubt. But how important has the dynamics of robustness and decay been in the history of life? I suggest two empirical approaches. The first approach applies experimental evolution of microbes. One may identify characters that perform some direct adaptive function, such as distinguishing correct from false signals or repairing damage. To the extent that those characters have a cost, robustness mechanisms that lessen the fitness sensitivity of the direct characters to perturbations should lead to the decay in the performance of the direct characters.

The second approach analyzes the phylogenetic history of robustness mechanisms that buffer against perturbations. The history of homeostasis provides a promising line of work. With each homeostatic innovation, how did the existing mechanisms that buffer against perturbations respond? Was there a net improvement in the total capacity of the system to deal with environmental perturbations, yet a decay in individual components of the overall homeostatic process? In general, does enhanced robustness lead to improved system performance but also the evolution of less costly, lower performance components?

## MATERIALS AND METHODS

The *Results* section includes the full methods for this study.

## ACKNOWLEDGMENTS

I was helped by discussions with David Krakauer.

## Author Contributions

Wrote the paper: SF.